\documentclass[conference, a4paper]{IEEEtran}
\IEEEoverridecommandlockouts
\usepackage{cite}
\usepackage{amsmath,amssymb,amsfonts}
\def\BibTeX{{\rm B\kern-.05em{\sc i\kern-.025em b}\kern-.08em T\kern-.1667em\lower.7ex\hbox{E}\kern-.125emX}}
\usepackage{algpseudocode}
\usepackage{algorithm}
\usepackage{graphicx}
\usepackage{textcomp}
\usepackage{xcolor}
\usepackage[left=1.62cm,right=1.62cm,top=1.9cm]{geometry}
\begin{document}
\IEEEoverridecommandlockouts
\IEEEpubid{\makebox[\columnwidth]{978-1-6654-6658-5/22/\$31.00~\copyright2022 IEEE \hfill}
\hspace{\columnsep}\makebox[\columnwidth]{ }}
\title{Addressing DAO Insider Attacks in IPv6-Based Low-Power and Lossy Networks}


\author{\IEEEauthorblockN{Sachin Kumar Verma}
\IEEEauthorblockA{\textit{Department of CSE} \\
\textit{PDPM IIITDM Jabalpur}, 
India \\
20mcs013@iiitdmj.ac.in}
\and
\IEEEauthorblockN{Abhishek Verma*}
\IEEEauthorblockA{\textit{Department of CSE} \\
\textit{PDPM IIITDM Jabalpur}, 
India \\
abhiverma@iiitdmj.ac.in}
\and
\IEEEauthorblockN{Avinash Chandra Pandey}
\IEEEauthorblockA{\textit{Department of CSE} \\
\textit{PDPM IIITDM Jabalpur}, 
India \\
avish.p@iiitdmj.ac.in}
}

\maketitle
\IEEEpubidadjcol
\begin{abstract}

Low-Power and Lossy Networks (LLNs) run on resource-constrained devices and play a key role in many Industrial Internet of Things and Cyber-Physical Systems based applications. But, achieving an energy-efficient routing in LLNs is a major challenge nowadays. This challenge is addressed by Routing Protocol for Low-power Lossy Networks (RPL), which is specified in RFC 6550 as a ``Proposed Standard" at present. In RPL, a client node uses Destination Advertisement Object (DAO) control messages to pass on the destination information towards the root node. An attacker may exploit the DAO sending mechanism of RPL to perform a DAO Insider attack in LLNs. In this paper, it is shown that an aggressive attacker can drastically degrade the network performance. To address DAO Insider attack, a lightweight defense solution is proposed. The proposed solution uses an early blacklisting strategy to significantly mitigate the attack and restore RPL performance. The proposed solution is implemented and tested on Cooja Simulator.    
\end{abstract}


\section{Introduction}
\begin{figure}
     \centering
     \includegraphics[width=.41\textwidth, trim={1cm 1cm 1cm 1cm}]{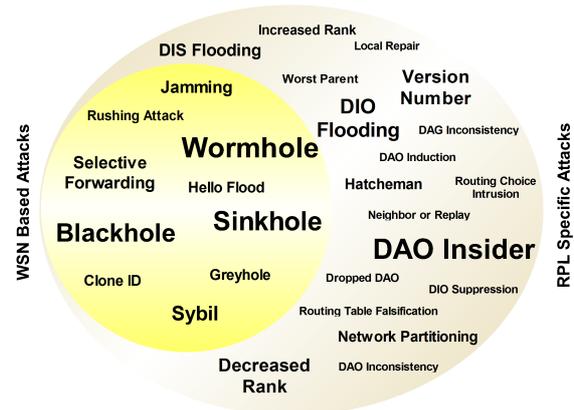}
     \caption{Routing attacks against RPL protocol}
     \label{fig:attacks}
\end{figure}

 The IoT\cite{ashton2009internet} has a large number of applications which make human life better. IoT applications like smart grid, smart healthcare, and smart agriculture require an infrastructure which has minimum implementation cost\cite{sethi2017internet} and also supports longer operation time. LLNs are the best for such applications as LLNs provide and infrastructure with a minimum implementation cost\cite{sobral2019routing} and has longer operation time. In LLNs, there are various security and privacy risks that may put user's security and privacy at risk. For example, auto-configuration, vulnerabilities of supporting devices and wireless communication may be explored by an attacker to access the confidential or private information of the users'. In addition, an attacker may target LLNs with Denial-of-Service attack and disturb the network's performance. To achieve minimum implementation cost and longer operation time, resource-constrained nodes are utilized . These nodes have very limited processing, storage, communication, and energy source capabilities. LLNs require an energy-efficient routing protocol for  network layer for supporting longer operation time. To address the problem of achieving energy-efficient routing in LLNs the IETF's Routing Over Low power and Lossy networks i.e. RoLL working group proposed a standard the RPL protocol. RPL is specified in RFC $ 6550 $\cite{rfc6550}. Although RPL solves major problems faced by LLNs, there are some LLNs characteristics(i.e., self-healing, self-organization, and resource-constrained behavior of nodes) which expose RPL to various outsider and insider attacks. Theses attacks may compromise user's privacy and security and limit the growth of IoT drastically. RPL has many discovered and undiscovered vulnerabilities which may be exploited by the attackers to compromise the network. An attacker may compromise resource-constrained devices and reprogram them to exploit vulnerable RPL features to disrupt the normal working of other legitimate nodes. In this manner, the attacker can continuously degrade the network’s overall performance. Fig. \ref{fig:attacks} indicates various attacks (WSN based and RPL specific) that can be performed on RPL based LLNs. Many of the attacks on RPL are very difficult to detect and mitigate. Fig. \ref{fig:attacks} shows some of the the most common attacks against RPL protocol.

One of the catastrophic attacks against RPL is known as DAO Insider attack. In this an attacker node can disrupt the network's performance by continuously sending DAO messages to its preferred parent node. As RPL does not has any inbuilt functionality to identify illegitimate control packets, therefore it becomes victim of such attack. To secure the network from DAO insider attacks, an Intrusion Detection System (IDS) is required. IDS may help RPL to detect the attack and mitigate it. Our contributions are summarized below:

\begin{enumerate}
\item A defense solution to address the DAO insider attack is proposed.
\item The effectiveness of our proposed solution is analyzed on the Cooja simulator.
\end{enumerate}

Further the paper is structured as follows. Section \ref{Sec:Overview of RPL protocol} overview's the RPL protocol. Section \ref{Sec:DAO insider attack} discusses the DAO Insider attack. Related works are discussed in Section \ref{Sec:related work}. Our proposed defense solution is described in Section \ref{Sec:proposed solution}. Performance evaluation of the proposed solution is depicted in Section \ref{Sec:performance evaluation}. Lastly, the Section \ref{Sec:conclusion} concludes the paper and indicates the future work.
 
\section{\textbf{Overview of RPL protocol}}\label{Sec:Overview of RPL protocol}
RPL is a proactive routing protocol based on distance vectors and source routing concepts. RPL is specified as a ``Proposed Standard" in RPL 6550 \cite{rfc6550}. RPL is considered as an energy-efficient protocol because it requires less energy to create and maintain network topology\cite{gaddour2012rpl}. It uses distance vector protocol for routing. RPL runs on top of IEEE $ 802.15.4 $ MAC. RPL forms Destination Oriented Directed Acyclic Graph (DODAG) based topology over LLN devices. DODAG is loop-free and tree-like structure in which root node is assumed as the destination for all the nodes. The network may be running several DODAGs at the particular instance of time which together are unidentified as an \textit{RPLInstance}. \textit{RPLInstance} is identified by a unique IPv6 address, i.e.,  \textit{RPLInstanceID}. In RPL, multiple \textit{RPLInstance} may be concurrently at the same time to support various services. Each LLN node is assigned a rank value which is a 16-bit integer and indicates the node's position relative to DODAG root node. RPL protocol defines a very strict rank rule. According to this rule, the rank of a nodes increases in a downward direction and decreases in an upward direction towards DODAG root. 
The concept of rank is used for following reasons: 

\begin{itemize}
    \item To recover the broken links.
    \item To differentiate between siblings and parents.
    \item To detect and resolve the routing loops.
    \item To create a relationship between parent and child.  
\end{itemize}

RPL supports four types of control messages, i.e., DODAG Information Solicitation (DIS), DODAG Information Object (DIO), Destination Advertisement Object (DAO), and Destination Advertisement Object Acknowledgment (DAO-ACK). 
RF defines Objective Function (OF) for rank calculation  \cite{lamaazi2020comprehensive}. OF is used to select optimal parent that that has shortest path towards DODAG root node. 
To reduce the number of control messages transmission RPL uses ``Trickle timer'' concept \cite{levis2011trickle}. 

\section{DAO Insider Attack}\label{Sec:DAO insider attack}
 To enable bi-directional communication, RPL uses DAO control messages. DAO messages are used to create downward paths so that DODAG root can route packets destined towards leaf nodes. DAOs are forwarded by each intermediate node that lies along the path between child node and DODAG root. Unicast DAO-ACK message is sent by a DAO Recipients that lie along the path. The standard RPL specification has not provided any information on when and how often these DAO messages must be transmitted. That is why different RPL implementations (i.e., ContikiRPL, OpenWSN, RIOT, Contiki-NG, OMNeT++, NetSim) choose different mechanisms to control DAO transmission rate. We have considered the most widely used RPL implementation, i.e., ContikiRPL in this paper. In ContikiRPL, DAO messages are transmitted using Trickle Timer. In RPL, DAO messages are unicast by the child node to parent node basically on three occasions:
 
 \begin{enumerate}
     \item When a node receives DIO message from a parent node.
     \item When a node changes its preferred parent.
     \item When a node detects some routing error.
 \end{enumerate}
 
\begin{figure*}
     \centering
     \includegraphics[width=.65\textwidth]{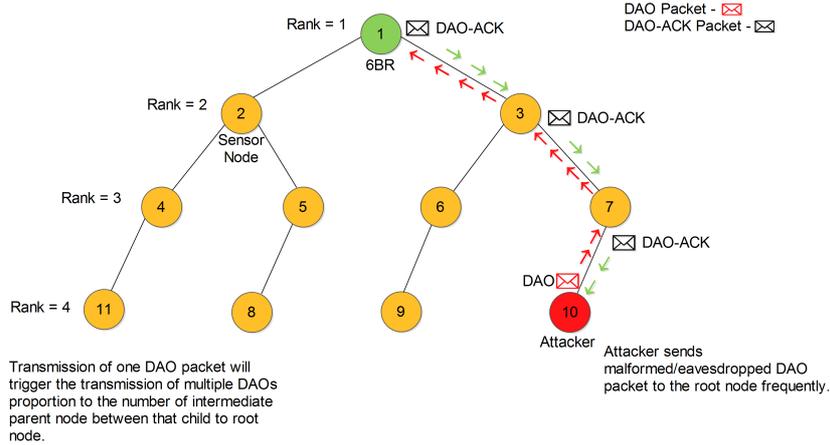}
     \caption{An illustration of DAO Insider Attack}
     \label{fig:dao_diagram}
 \end{figure*}

 An important point related to DAO messages is that when a child node sends a DAO message with DODAG root as a destination, in response to a single DAO transmission multiple DAO messages are generated by intermediate nodes that are present along the path. Consider a path from child node to root node that consists of $n$ intermediate nodes, then the total number of DAO messages that are transmitted along the path is equal to $n$, as shown in Fig. \ref{fig:dao_diagram}. An attacker node may exploit this feature to disrupt the normal network's performance by simply transmitting malformed or eavesdropped DAO message frequently to its preferred parent node. The best case scenario for an attacker will be to launch the attack from the edge of the network as this will increase the control packet overhead in terms of DAO messages. DAO Insider attack significantly decreases the PDR  (packet delivery ratio), increases AE2ED (average en-to-end delay) and avearge power consumption of the network. There are multiple ways to launch the DAO Insider attack. One way is to is send malformed DAO packets to the root node (i.e., insider attack). Another way is to transmit an eavesdropped DAO captured from legitimate node (i.e., outsider attack). In Fig. \ref{fig:dao_diagram} it is shown that the attacker with Node\_Id 10 is repeatedly transmitting the DAO message to the preferred parent node, i.e., Node\_Id 7. All intermediate nodes forward the DAO message to their parent until it received by the root node. 

\section{Related Work}\label{Sec:related work}
\textcolor{black}{Sheibani \textit{et al.} \cite{sheibani2022lightweight} proposed an algorithm for mitigating Dropped DAO (DDAO) attack. They used a watchdog approach to monitor the forwarding behaviour of its parent.} Raza \textit{et al}. \cite{raza2013svelte} suggested a real-time IDS called SVELTE which is based on Contiki platform. SVELTE detects three types of attacks, i.e., Sinkhole, Selective Forwarding, and Spoofing. It uses three different procedures to detect attack in real-time: (1) collects traffic information, (2) identifies intrusion, (3) provides a small distributed firewall for blocking illegitimate traffic coming from outsider networks. Verma \textit{et al}. \cite{verma2020security} carried out a detailed survey on various existing attacks and countermeasures for RPL. Mayzaud \textit{et al}. \cite{mayzaud2017distributed} proposed a distributed monitoring algorithm to secure RPL from version number attacks. 
In \cite{ioulianou2018signature}, the focused on  designing an IDS to protect the network from outsider attackers. They proposed a signature-based intrusion detection approach to secure the network from version number modification and ``Hello” flooding attacks. \textcolor{black}{In \cite{cakir2020rpl}, an attack classification model based on Gated Recurrent Unit network is developed for identification of ``Hello Flooding" attack.} Ghaleb \textit{et al}. \cite{ghaleb2018addressing} proposed and addressed the DAO Insider attack. The authors implemented a defense mechanism named SecRPL to secure the LLNs. 
Verma \textit{et al}. \cite{verma2020mitigation} proposed a lightweight security scheme for the defending RPL against DIS flooding attacks. They analyzed the network and put a safety threshold on the RPL protocol. \textcolor{black}{In this\cite{8765272} paper Farzaneh \textit{et al.} proposed an anomaly based IDS based on threshold values for detection of attacks in RPL.} Ariehrour \textit{et al}. \cite{airehrour2019sectrust} proposed SecTrust-RPL solution to secure RPL against Sybil and rank attacks. AN IDS named SIEWE is proposed by Patel \textit{et al}. by Patel \textit{et al.} \cite{patel2019blackhole}.  In \cite{guo2021lightweight}, the authors proposed a lightweight mechanism that adjusts thresholds value to detect and mitigate DIS attacks. From the literature, we identified that various RPL based attacks have been countered using different types of security solutions. As far as the literature is concerned there is only one solution for defending DAO Insider attacks \cite{ghaleb2018addressing}, this leave a lot of scope. In this paper we have addressed DAO Insider attack using Blacklisting technique.

\section{Proposed Solution}\label{Sec:proposed solution}
The proposed defense solution is based on the idea of analyzing node's behavior to identify whether it is legitimate or illegitimate. We performed multiple experiments considering different non-attack and attack scenarios to analyze the illegitimate node behavior. The behavior of the node is analyzed in form of the number of DAO messages being transmitted and received by the nodes across the network. With a detailed analysis, we come to a conclusion that each node in RPL based LLNs receives and transmits similar number of DAOs messages in the network under non-attack scenarios. Whereas, in case of attack, victim node receives large amount of DAOs from a malicious node as compared to neighbor legitimate nodes. To address DAO Insider attack, we proposed a defense solution that puts limits on the the number of DAOs messages sent by any child node. The key idea is to distinguish between original attacker node and victim node in order to minimize false positives. The proposed solution is based on distributed detection strategy in which every individual node maintains two tables, i.e., a neighbor table for storing information about neighbors, a blacklist table to store information about blacklisted or attacker nodes. The usage of blacklist table helps in energy saving because attack is mitigated quickly without additional processing of illegitimate DAO packet. A threshold, i.e., DAO\_recv\_threshold is used to put a cap on the maximum allowed DAO transmissions by any child node. The value of DAO\_recv\_threshold is chosen based on the analysis of multiple non-attack scenarios. The detection algorithm starts with the initialization of DAO\_recv\_threshold, Neighbor\_Table, and Blacklist\_Table. The parent node, upon receiving a DAO message from a child node or DAO sender checks whether the DAO sender's address is already present in the Blacklist\_Table or not. If DAO sender's address matches with any blacklisted node's address, this means that parent had already detected that DAO sender as an attacker node earlier, and it simply discards received DAO message without any further processing. This not only saves energy of nodes but also helps in quick mitigation of attack. In case the DAO sender's address is not present in Blacklist\_Table, then the algorithm starts checking the Neighbor\_Table to find out the DAO sender's address. If DAO sender's address is not present in the Neighbor\_Table, then it means that DAO sender is a new child node which has sent the DAO message first time. Then, a new node entry in the Neighbor\_Table is created and DAO sender's information is added to the Neighbor\_Table. In this case the Neighbor\_Table stores three values:

\begin{enumerate}
    \item DAO\_sender\_address(Node[source\_id])
    \item Child's Global address or DAO\_Prefix(Node[global\_id])
    \item Child's DAO\_counter
\end{enumerate}

Based on these entries, the detection algorithm decides whether a DAO sender node is an attacker or not. It is important to note that whenever a node generates a DAO message, it also transfers the global ID in the DAO message. In RPL, DAO sender's global ID is represented as the DAO prefix. In our solution, we use the DAO prefix to increment the DAO counter value (i.e., DAO\_count). Whenever any parent node receives a DAO message from its child node there are two cases which are handled differently. In first case, if  DAO sender or child is the DAO originator (i.e., DAO\_Prefix equals child's global\_id), then DAO\_count value corresponding to that child node is incremented, and the DAO message is forwarded. In second case, when the child is not the DAO originator (i.e., DAO\_Prefix not equals child's global\_id), the value of DAO\_count is not incremented, and DAO message is forwarded to the preferred parent. With this approach, the algorithm detects attacker node present in the network without blacklisting legitimate nodes. If any node is sending a lot of DAO message, then the parent of that child node will increment the DAO\_Counter corresponding to that child node. After reaching the DAO\_recv\_threshold, the parent blocks the abnormally behaving child and add its information in the Blacklist\_Table (i.e., blacklisting). \textcolor{black}{The main benefit of this approach is that it does not involve usage of any resource consuming methods like encryption, decryption, hashing, or key management. The detection logic simply puts thresholds of RPL parameters which makes it lightweight and suitable for LLNs. Pseudocode of the proposed solution is depicted in Algorithm \ref{Algo}.}

\begin{algorithm}[!h]
\small
\caption{Pseudocode of proposed solution}
\label{Algo}
\begin{algorithmic}[1]
\Procedure{Initialization}{}
    \State {set DAO\_recv\_threshold}
    \State {create empty Neighbor\_Table} \Comment{To create a neighbor table on node start}
    \State {create empty Blacklist\_Table} \Comment{To create a blacklist table on node start}
\EndProcedure
\Procedure{On\_DAO\_Receive}{}
\If {(DAO\_sender\_address \textbf{is present in} Blacklist\_{Table})}
\State \textbf{return} \Comment{In case the sender node was already blacklisted}
\EndIf
\For{Each Node \textbf{in} Neighbor\_Table}
\If {(DAO\_sender\_address \textbf{equals} Node.source\_id)}
    \If {(DAO\_Prefix \textbf{equals} Node.global\_id)}
        \If {(Node.DAO\_count \textbf{is less than} DAO\_recv\_threshold)}
        \State Node.DAO\_count++
        \State Forward DAO to preferred parent
        \Else
        \State Add DAO sender \textbf{in} Blacklist\_{Table}
        \EndIf
    
     \Else
    \State Forward DAO to preferred parent
   \EndIf
 \Else
\State Add new DAO sender's information \textbf{in} Neighbor\_Table
\EndIf      
      \EndFor

\EndProcedure
\end{algorithmic}
\end{algorithm}

\section{Performance Evaluation}\label{Sec:performance evaluation}

We implemented our proposed defense solution in Contiki \cite{dunkels2004contiki} which is one of the widely used embedded operating system for resource constrained nodes.  The popular ContikiRPL is modified and the proposed solution is integrated with it. The performance of the proposed solution is evaluated on Cooja simulator \cite{osterlind2006cross}. Further part of this section provides the details of experimental setup, performance indicators, and experimental results.

 \begin{figure}
     \centering
     \includegraphics[width=0.4\textwidth]{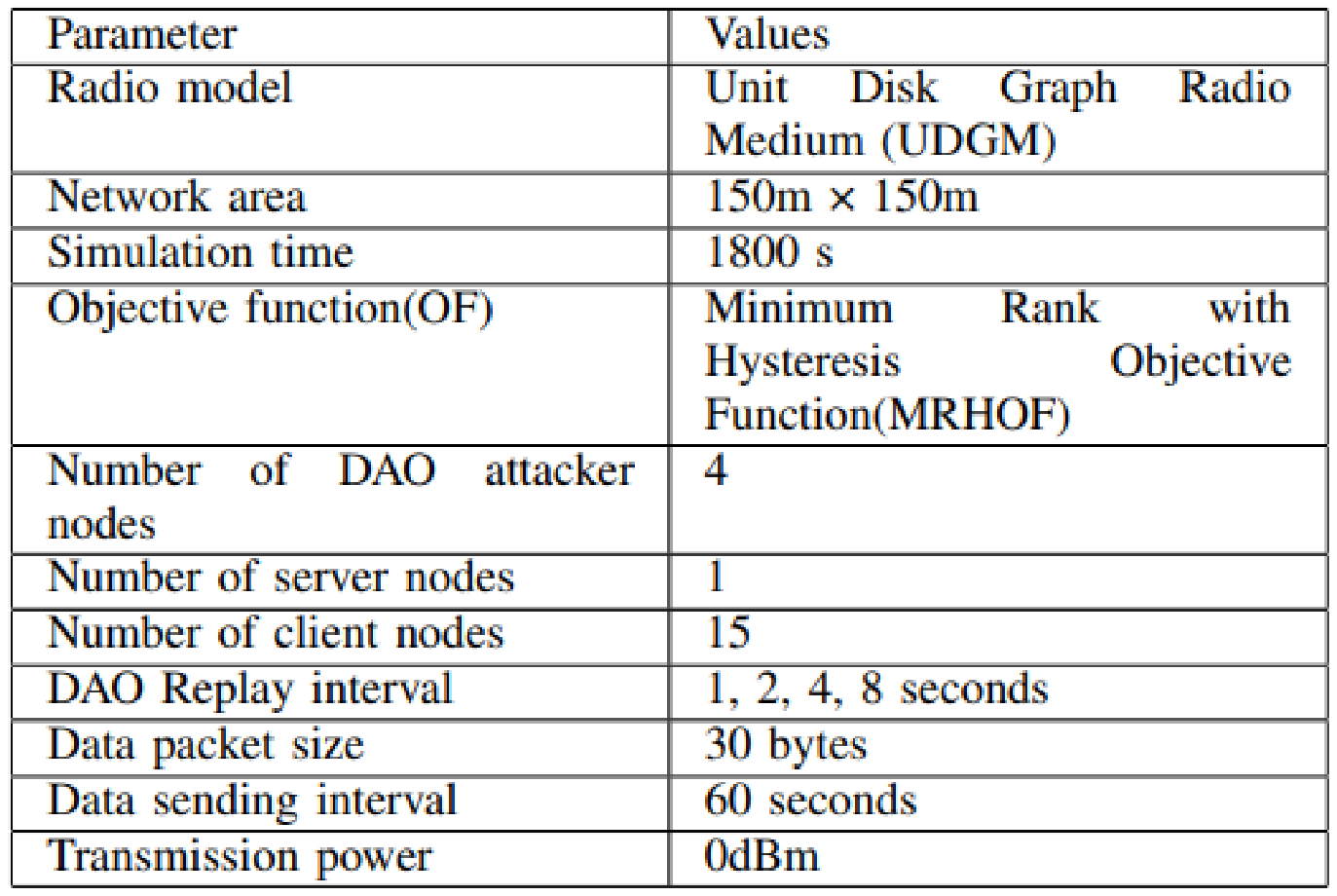}
     \caption{Simulation Parameters}
     \label{fig:simp}
 \end{figure}

        


\subsection{Experimental Setup}

The proposed solution is implemented by modifying the core files of ContikiRPL. We performed the experiments for evaluation of proposed solution on Cooja Simulator. Z1 mote platform is used for running Contiki. The simulation parameters mentioned in Fig \ref{fig:simp} are considered for experiments. In all the experiments, the Unit Disk Graph Radio Medium (UDGM) is considered. To mount the DAO insider attack, an attacker node can compromise the legitimate node and reprogrammed it to capture the DAO message and then transmit the captured DAO message in a fixed time of interval. The DAO attack is launched after receiving a DIO message from any parent node. The detection approach of the proposed solution is activated upon network initialization. The mean values of PDR and AE2ED have been used for analysis to eliminate the effect the biased results. We performed 10 independent experiments with different random seed values and computed the errors at a 95 percent confidence interval.

\subsection{Performance Indicators}
\begin{enumerate}
\item Packet Delivery Ratio (PDR): Represents ratio of the total amount of data packets received to the total amount of data packets sent by any node to the DODAG root node.
    
\item Average End-to-End Delay (AE2ED): The average time taken to deliver all the data packet from source to DODAG node. 
    
    
    \item \textcolor{black}{Throughput: It indicates the amount of data moved successfully from sender to receiver in a given period of time. It is expressed in terms of bits per second (bps).}
\item Implementation Overhead: It represents total RAM and ROM usage by the proposed solution over resource constraints nodes.
\end{enumerate}

\subsection{Simulation Results}

We have considered three cases for making comparisons, i.e., $RPL$, $RPL_{Under Attack}$, and $RPL_{Secure}$. Where, $RPL$ represents standard RPL without defense mechanism implemented on it, $RPL_{Under Attack}$ is the scenario in which standard $RPL$ is under attack, and $RPL_{Secure}$ represents the secure version of standard $RPL$ which has our defense solution incorporated in it. In this section, the simulation results are discussed.

\subsection{Impact on PDR}
\begin{figure}[h]
    \centering
    \includegraphics[width=.45\textwidth]{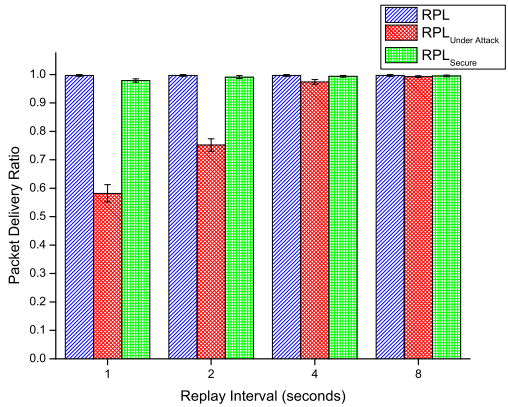}
    \caption{PDR values obtained in different scenarios}
    \label{fig:PDR}
\end{figure}

Fig. \ref{fig:PDR} represents the impact of PDR on $RPL$, $RPL_{Under Attack}$, $RPL_{Secure}$. It has been observed from the figure that the attacker lowers the network's performance. Under  $RPL_{Under Attack}$ scenario, the attacker node is programmed to transmit a large number of DAO messages to the preferred parent node. The attacker node increases the control packet overhead of the network. upon receiving a DAO message a parent must processes all DAOs and sends acknowledgement in DAO-ACK message to the DAO sender node. In $RPL_{Under Attack}$ case the processing overhead increases drastically which consequently leads to data packet loss. Fig. \ref{fig:PDR} clearly indicates how the PDR is affected in $RPL_{Under Attack}$ scenario. Moreover, it can also be analyzed that the aggressive DAO attacker (i.e., attacker sending DAO at 1, 2 second replay interval) have high impact on PDR as compared to non-aggressive attackers ((i.e., attacker sending DAO at 4, 8 second replay interval)). In case of $RPL_{Secure}$, whenever a parent node receives DAO messages greater than threshold value, the parent node will block the DAO sender node and discard the further received DAOs from that node. $RPL_{Secure}$ is able to improve the network performance and reduces the impact of attack. The effectiveness of proposed solution is clearly visible from the values achieved in case of $RPL_{Secure}$ as shown in the Fig. \ref{fig:PDR}.

\subsection{Impact on AE2ED}
\begin{figure}[h]
    \centering
    \includegraphics[width=.45\textwidth, trim={0cm 1.8cm 0cm 0cm}]{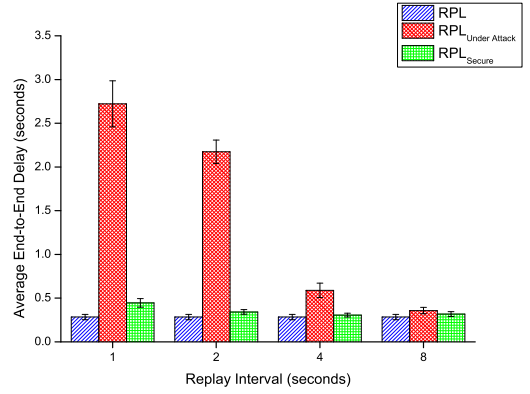}
    \caption{AE2ED values obtained in different scenarios}
    \label{fig:AE2ED}
\end{figure}

The impact of AE2ED in different scenarios ($RPL$, $RPL_{Under Attack}$, $RPL_{Secure}$) is indicated in Fig. \ref{fig:AE2ED}. It can be observed that AE2ED is severely affected under attack scenario as compared to $RPL$. The reason is that the parent node receives a lot of DAO messages from the attacker node and this keeps them busy. Busy parent nodes take a lot of time to process data packets, therefore AE2ED increases. Like, PDR results in this case also, it can also be analyzed that the aggressive DAO attacker have major impact on AE2ED of the network as compared to non-aggressive attackers . Our proposed solution ($RPL_{Secure}$) is able to decrease the impact of attack and this is clearly visible in Fig. \ref{fig:AE2ED}. This is because the proposed solution discards malicious DAOs, which consequently reduces processing time of data packets.

\subsection{Impact on Throughput}
\begin{figure}[h]
    \centering
    \includegraphics[width=.45\textwidth, trim={0cm 1.8cm 0cm 0cm}]{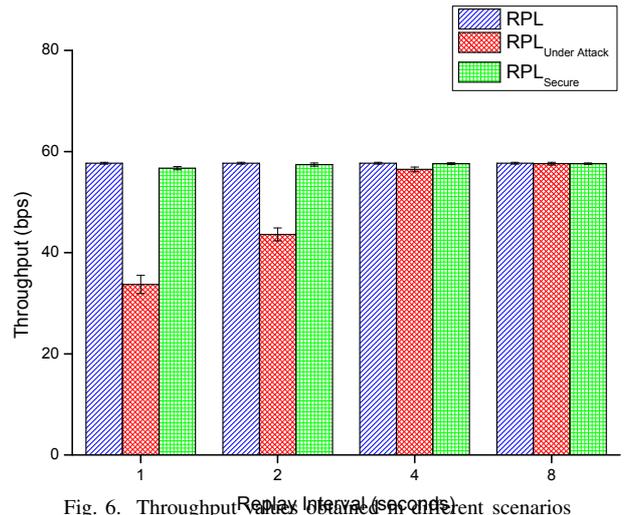}
    \caption{Throughput values obtained in different scenarios}
    \label{fig:TP}
\end{figure}

\textcolor{black}{It can be observed from the results shown in Fig. \ref{fig:TP} that $RPL_{Secure}$ is able to improve the throughout (data packet bits delivered) of the network which is decreased due to effect of attack ($RPL_{Under Attack}$). The proposed solution reduces the effect of DAO insider attack and therefore the number of data packets successfully delivered are increases which consequently increases throughput of the network.}

\subsection{Implementation Overhead}

\begin{figure}[h]
    \centering
    \includegraphics[width=.5\textwidth, trim={0cm 1.8cm 0cm 0cm}]{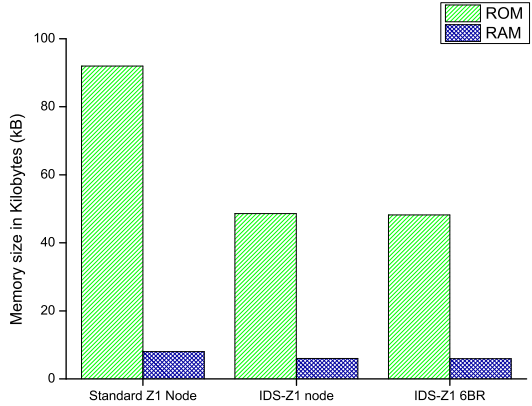}
    \caption{Memory requirements of proposed solution}
    \label{fig:memory}
\end{figure}

Fig. \ref{fig:memory} shows the memory requirements of proposed defense solution. The proposed solution requires very little amount of memory hence it becomes a lightweight defense solution. The standard Z1 motes have 92 KB of ROM, and 8 KB of RAM. Fig. \ref{fig:memory} shows that Contiki with our proposed solution implemented on it easily fits into Z1 motes without imposing significant overhead. Thus, the implementation overhead of proposed solution makes it lightweight solution.

\subsection{Time complexity of Proposed Approach}

\begin{itemize}
    \item \textcolor{black}{The time complexity of the INITIALIZATION procedure is $O(1)$ as it defines the neighbor and blacklist table.}
    \item \textcolor{black}{ON\_DAO\_Receive procedure explores the blacklist table to determine whether unauthorized senders are already blacklisted or not. If the size of the blacklist table is $B_t$ and unauthorized senders are present in the blacklist table, then the time taken to explore the entire blacklist table will be $O(B_t)$. The neighbor table is explored to discover the unauthorized senders if the senders are not present in the blacklist table. After identifying the unauthorized senders, it is added to the blacklist table. If the size of the neighbor table is $N_t$, then the time complexity to discover and add an unauthorized sender to the blacklist table will be $O(B_t)+O(N_t)$ because the neighbor table is explored after examining the entire blacklist table.}
\end{itemize}
\textcolor{black}{The time complexity of the proposed approach will be $O(B_t)+O(N_t)+O(1)$, i.e., $O(B_t)+O(N_t)$, since the time taken by the initialization procedure is $O(1)$, and ON\_DAO\_Receive procedure is $O(B_t)+O(N_t)$.}

\section{Conclusion and Future Scope}\label{Sec:conclusion}


\textcolor{black}{In this paper, we have proposed a lightweight defense solution to address DAO Insider attacks in LLNs.  The experimental results indicate that our proposed solution effectively detects and mitigates the attack while taking care of the resource nature of LLN nodes. In future, we aim to test our proposed approach in dynamic network scenarios and perform tested experiments.}  

\bibliographystyle{IEEEtran}
\bibliography{ref}
\end{document}